\documentclass[usegraphicx,usenatbib]{mn2e}
\usepackage{psfig}

\renewcommand\[{\begin{equation}}
\renewcommand\]{\end{equation}}

\def\kpc{\,{\rm kpc}}

\def\erg{\,{\rm erg}}\def\ergs{\,{\rm erg\,s}^{-1}}
\def\s{\,{\rm s}}

\def\yr{\,{\rm yr}}\def\Myr{\,{\rm Myr}}

\def\msun{{\rm M_{\odot}}}
\def\kms{\,{\rm km}\, {\rm s^{-1}}}

\def\i{\relax\ifmmode{\rm i}\else\char16\fi}

\def\lesssim{{_ <\atop{^\sim}}}

\def\fracj#1#2{{\textstyle{#1\over#2}}}

\def\lesssim{\mathrel{\hbox{\rlap{\hbox{\lower4pt\hbox{$\sim$}}}\hbox{$<$}}}}
\def\gtrsim{\mathrel{\hbox{\rlap{\hbox{\lower4pt\hbox{$\sim$}}}\hbox{$>$}}}}

\def\apj{ApJ}

\def\mnras{MNRAS}

\def\apjl{ApJ}
\def\aap{A\&A}

\begin{document}

   \title[Stability of cooling flows]
   {Structural stability of cooling flows}

   \author[Omma \& Binney]
          {Henrik Omma, James Binney\\
		  Department of Physics, Theoretical Physics, 1 Keble Road, Oxford OX1 3NP\\
          }

\date{Received ...; accepted ...}

   \maketitle

\begin{abstract}
Three-dimensional hydrodynamical  simulations are used to investigate the
structural stability of cooling flows that are episodically heated by jets
from a central AGN.  The radial profile of energy deposition is controlled
by (a) the power of the jets, and (b) the pre-outburst density profile. A
delay in the ignition of the jets causes more powerful jets to impact on a
more centrally concentrated medium. The net effect is a sufficient increase
in the central concentration of energy deposition to cause the post-outburst
density profile to be less centrally concentrated than that of an identical
cluster in which the outburst happened earlier and was weaker. These results suggest
that the density profiles of cooling flows oscillate around an attracting
profile, thus explaining why  cooling flows are observed to have similar
density profiles. The possibility is raised that powerful FR II systems are
ones in which this feedback mechanism has broken down and a runaway growth
of the source parameters  has occurred.
\end{abstract}

\begin{keywords}
cooling flows -- X-rays: galaxies: clusters -- galaxies: jets -- hydrodynamics
\end{keywords}

\section{Introduction}

Data from the {\it Chandra\/} and {\it XMM-Newton\/} X-ray observatories
have established beyond reasonable doubt that radio sources associated with
the central black holes of X-ray emitting clusters are heating the
surrounding `cooling flows' at rates comparable to the radiative cooling
rate from within the cooling radius (where the cooling time is equal to the
Hubble time).  The key observation is that several clusters contain pairs of
`cavities' -- regions of depressed X-ray surface brightness that in some
cases coincide with regions of enhanced brightness in the radio continuum
\citep{BohringerEtal93,ChurazovEtal00,McNamaraEtal00,McNamaraEtal01,BlantonEtal01}.
It is clear that the observed radio jets are inflating pairs of cavities in
the thermal X-ray emitting plasma, and that synchrotron emission from the
cavities accounts for the enhanced radio-frequency brightness.

Once a cavity is created, it must rise buoyantly
\citep{GullNorthover,ChurazovEtal01,QuilisEtal01,BruggenKaiser01,BruggenKaiser02,BruggenKCE02},
and it may rise even faster if it is endowed at birth with significant
momentum as well as energy \citep{Ommaetal}. Simulations of cavity dynamics
all affirm that cavities rise at speeds comparable to the sound speed in the
X-ray emitting gas, $\sim1000\kms$ in a typical cluster. In Perseus
\citep{Fabianetal00}, Abel 2597 \citep{McNamaraEtal01} and Abel 4059
\citep{HeinzCRB} more than one pair of cavities is detected. The distance
from the cluster centre of the outer cavities, combined with the speed at
which cavities rise, yields an estimate $\tau\sim50\Myr$ of the time
between successive episodes of cavity inflation.

The crudest estimate of the energy injected during cavity inflation is
$\fracj52PV$ if the plasma in the cavity is subrelativistic, or $4PV$ in the
contrary case. Since cavities must be irreversibly inflated, these estimates
for reversible inflation are underestimates. Binney (2003) gives estimates
for five clusters of the ratio $\tau$ of the estimate $3PV$ of the energy
injected during the inflation of observed cavities to the X-ray luminosity.
The values of $\tau$, which are significantly uncertain, range from $30\Myr$
to $120\Myr$, implying that an outburst needs to occur each $50$ to
$100\Myr$ to maintain a balance between radiative losses and heating by the
radio source. It is striking that this estimate of $\tau$ from the
hypothesis that heating balances cooling coincides to within the
uncertainties with the direct estimate of $\tau$ yielded by clusters with
two pairs of cavities.  Arguments from synchrotron aging have long implied
that more powerful radio sources have lifetimes of the order $100\Myr$
\citep{Pedlar90}, but here the conclusion is slightly different: individual
AGN outbursts might last a significantly shorter time that $100\Myr$, but
the interval {\it between\/} outbursts is of this order. 

It is to be expected that a drop in the central temperature of cooling-flow
gas will increase the accretion rate of the embedded black hole, and thus
the rate at which the radio source heats the flow (Tabor
\& Binney 1993). Consequently, the possibility exists for a steady-state
balance between heating and cooling, such as occurs in the cores of
main-sequence stars. However, there are many observational indications that
the luminosities of radio sources are unsteady -- for example the
existence of cavities and of powerful FR II radio galaxies, which in
$100\,\Myr$ can pump as much energy into the intracluster medium as it can
radiate in a Hubble time. We need a better understanding of the cause and
extent of this unsteadiness.

Another area in which we urgently need a better understanding is in the radial
distribution of heat input by the radio source. Tabor \& Binney (1993)
assumed that heat was applied at very small radii, leading to the
development of an adiabatic core. Recent X-ray data show that the specific
entropy decreases inwards to the smallest radii probed (Kaiser \& Binney
2003). Such a situation arises naturally when the radio source heats the
X-ray gas with a collimated outflow, because heating then occurs where the
outflow energy thermalizes, which is expected to extend over a significant
range in radius (Binney \& Tabor 1995).

The radial distribution of energy input by the radio source depends on two
factors: (i) the power and degree of collimation of the outflow, and (ii)
the radial density profile of the intracluster gas. The more powerful and
better collimated the outflow is, the further out it will go before it
disrupts. Hence powerful, strongly collimated outflows, such as those of FR
II radio galaxies, deposit most of their energy at large radii, and probably
outside the cooling radius
\citep{ReynoldsHeinzBegelman01,ReynoldsHeinzBegelman02,BassonA}.
The more steeply the density of the X-ray gas increases towards the cluster
centre, the more readily the gas can disrupt a given jet, and thus the more
centrally concentrated the energy injected into the ICM will be.

These considerations raise a question about the stability of cooling flows.
Consider two initially identical cooling flows that are both experiencing
periods of nuclear quiescence. In flow 1 the radio source switches on at
time $t_1$, while that in flow 2 switches on at a later time $t_2$. On
account of the additional period $t_2-t_1$ of cooling experienced by flow 2,
the jet in this system will impact on a more centrally concentrated
distribution of gas than that impacted by the jet of flow 1. On the other
hand, if heating and cooling are to be in statistical balance in both flows,
the luminosity of the jet in flow 2 must be larger than that in flow 1. This
greater luminosity will tend to make the energy deposition less centrally
concentrated. 

If the effect on the radial distribution of energy input from
enhanced luminosity is greater than that from increased central
concentration of the X-ray gas, the post-outburst density profile of flow 2
will be more concentrated than that of flow 1.  A runaway situation will
then arise, in which ever more powerful outbursts, driven by a steadily
rising central gas density, deposit their energy at larger and larger radii.
FR II sources might be the cataclysmic end points of such a runaway. 

In the contrary case, the dominating factor is the tendency of the steeper
density profile of the pre-outburst gas to concentrate injected energy 
at small radii, and in flow 2 the post-outburst gas is
no more centrally concentrated than that in flow 1. In this case the density
profiles of both flows will tend to fluctuate around some stable density
profile of the X-ray emitting gas. The fact that many cooling flows have
density profiles that differ mainly in an overall scaling, even though their
central cooling times are of order a thirtieth of the Hubble time, suggests
that there is a significant range of parameters in which such stability is
possible.

In this paper we use simulations to
investigate the existence of such a stable configuration.

\begin{table}
\caption{Parameters of the simulations\label{tab1}}
\begin{tabular}{l|ccccc}
&$\dot m_{\rm jet}$&$v_{\rm jet}$&$P_{\rm jet}$&$\Delta t$&$\Delta E$\\
&$\msun\yr^{-1}$&$10^3\kms$&$10^{44}\erg\s^{-1}$&$\Myr$&$10^{59}\erg$\\
\hline
S1&1&28&5&25&4\\
S2&1&40&10&25&8\\
S3&1&28&5&50&8\\
S4&1&28&5&$2\times25$&8\\
S5&1&40&10&25&8\\
\hline
\end{tabular}
\end{table}
\section{The simulations}

We have used the Eulerian hydrocode ENZO \citep{BryanNorman97,Bryan99} to make a
suite of five simulations of cooling-flow evolution. The simulations are
fully three-dimensional and use an adaptive mesh to achieve an effective
(maximum) resolution of $0.61\kpc$. The simulations employ the PPM Riemann
solver \citep{ColellaWoodwardPPM84}. The computational volume is a box
$628\kpc$ on a side, with periodic boundary conditions. The twin jets were
imposed by the algorithm described in \cite{Ommaetal}. Table~\ref{tab1}
lists the parameters of the simulations. The quantities $\dot m_{\rm jet}$,
$v_{\rm jet}$ and $P_{\rm jet}$ quantify the rate at which mass, momentum
and energy are injected at the base of the jet through the formulae
$\dot p=\dot m_{\rm jet} v_{\rm jet}$ and $P_{\rm jet}=\fracj12\dot m_{\rm jet}v_{\rm
jet}^2$. 

The initial conditions are for gas in hydrostatic equilibrium in an NFW
\citep{NFW97} gravitational potential. The gas cools
radiatively throughout the simulation. The central cooling time is initially
$380\Myr$, so cooling steepens the density profile quite rapidly.  After
~$250\Myr$ each simulation has a distinctly cooled, high-density core.

In Simulation 1 the jets fire after $262\Myr$ of cooling. They have a total
power of $5\times10^{44}\ergs$ and run for $25\Myr$, during which time they
inject $4\times10^{59}\erg$. The jets in  Simulations 2 to 4 fire
after $300\Myr$ of cooling, by which time an extra $4\times10^{59}\erg$ has
been lost to radiation, and they inject $8\times10^{59}\erg$. Thus the later
ignition of the jets in Simulations 2 to 4 is compensated for by enhanced
energy injection. 

\begin{figure}
\centerline{\psfig{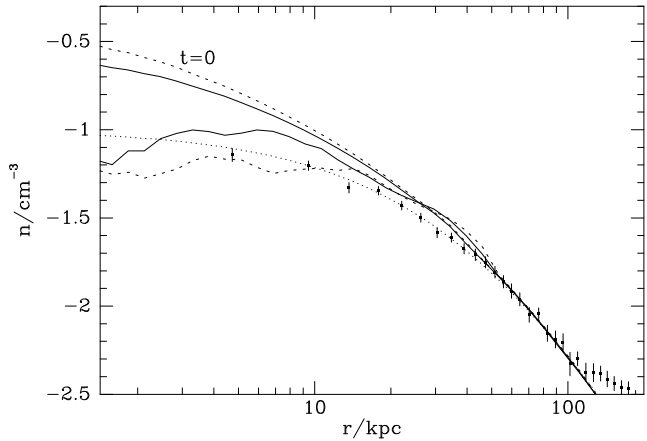}}%
\caption{Dotted curve: initial density profile of all simulations. Data
points: density in Hydra from David et al.\ \citep{DavidEtal00}.  Curves
labelled $t=0$: densities
after cooling and immediately before jet ignition in Simulation 1 (full
curve) and Simulation 2 (dashed curve). Bottom  curves show
spherically averaged density profiles $42\Myr$
after ignition in Simulation 1 (full) and Simulation 2 (dashed).
\label{densfig}}
\end{figure}

Simulations 2 to 4 differ in the pattern of their outbursts.  The jets in
Simulation 2 are twice as powerful as those in Simulation 1 and run for the
same time ($25\Myr$). The jets in Simulation 3 have the same power as those
in Simulation 1 but fire for twice as long ($50\Myr$). The jets in
Simulation 4 have the same power but fire for two $25\Myr$ intervals,
separated by a quiescent interlude $25\Myr$ long.

In Simulation 5 the jets fire at the same early time as in Simulation 1, but
with the power level and duration that are characteristic of Simulation 2.

In Fig.~\ref{densfig} the dotted curve shows the density profile of the
cluster gas at the start of all simulations. The data points show the
density in the Hydra cluster as deduced by \cite{DavidEtal00}. The full
curve labelled $t=0$ shows the density profile at the ignition of the jets
in Simulations 1 and 5, while the dashed curve labelled $t=0$ shows the
density profile at the ignition of the jets in Simulations 2 to 4. The
effect on the density profile of $\sim300\Myr$ of passive cooling is
evident.  The bottom full curve shows the spherically averaged density
profiles $42\Myr$ after the firing of the jet in Simulation 1, while the
bottom dashed curve shows the same data for Simulation 2. At that time,
$17\Myr$ after the jets extinguished, the curves are quite similar to the
initial profile and the data. Thus in both simulations the injected energy
has effectively reversed the effect of $300\Myr$ of cooling. 

Most crucially, the dashed curve of Simulation 2 now lies {\it below\/} the
full curve, implying that the system that cools for longer and has the most
centrally concentrated density profile when its jets ignite, ends up with
the {\it less\/} centrally concentrated profile.  The density profiles at
times later than those shown in Fig.~\ref{densfig} confirm that the greater
central concentration of Simulation 1 at $t=42\Myr$ is not an aberration:
the profile for Simulation 1 remains on top of that of Simulation 2, and
moves upwards faster. Consequently, when the profiles are next similar to
those labelled $t=0$ in Fig.~\ref{densfig}, we can expect Simulation 1 to be
the scene of the more energetic outburst slamming into the more centrally
concentrated ICM. When the dust settles after this second outburst, the
profile of Simulation 2 will  be the more centrally concentrated and
the pair of simulations will have come full cycle.  Hence these simulations
strongly support the proposition that the density profiles of cooling-flow
clusters oscillate around an attracting profile.

\begin{figure}
\centerline{\psfig{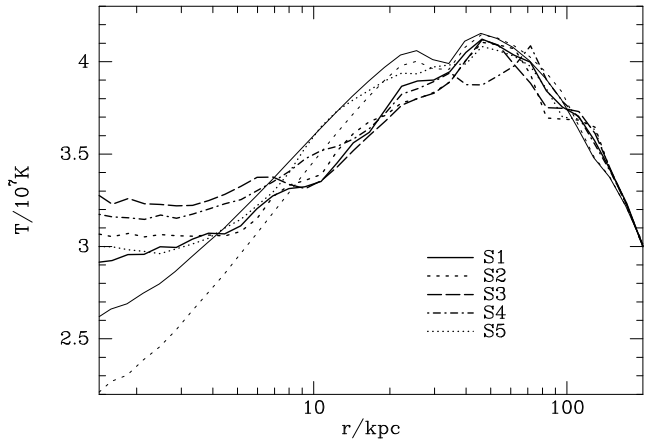}}
\caption{Light curves: temperature profiles when the jets ignite; full curve
Simulations 1 and 5; dashed curve Simulations 2 to 4. Heavy curves:
spherically averaged temperature profiles $120\Myr$ after jet
ignition.\label{tempfig}}
\end{figure}

Simulations 3 to 5 probe the sensitivity of a cooling flow's response to the
temporal pattern of energy injection. In Fig.~\ref{tempfig} the light curves
show the temperature profiles just before jet ignition: the full curve
applies to simulations 1 and 5 that experience early ignition, while the
dashed curve applies to Simulations 2 to 4. The full curves show the
spherically averaged temperature profiles $120\Myr$ after ignition. Interior
to $r\sim10\kpc$ the dotted curve of Simulation 5 pretty much tracks the
heavy full curve of Simulation 1, showing that the extra energy injected in
Simulation 5 does not appear near the centre, but in the radial range 10 to
$30\kpc$. By contrast, at the centre the heavy dashed curve of Simulation 2
lies distinctly above the curves for Simulations 1 and 5. Since the jets in
Simulations 2 and 5 are identical in power and duration, the higher central
temperature of Simulation 2 at $120\Myr$ must be due to the injected energy
being concentrated in the centre by the more centrally concentrated
pre-ignition density profile.

Fig.~\ref{tempfig} shows that the weaker, longer-lasting jets of Simulations
3 (long-dashed curve) and 4 (dash-dot curve) produce higher central
temperatures than the short-lived powerful jets of Simulation 2.  Thus
energy injected by a weaker jet is more centrally concentrated than energy
injected by a stronger jet since in a given intracluster medium a weaker jet
will disrupt at a smaller radius. Another relevant factor is revealed by
examination of plots of specific entropy: in part jets
heat the core by entraining and carrying upward some of the coldest gas.
In Simulation 2 much of what is initially lifted out of the core falls
straight back after the jets have died. The weaker but longer lasting jets
of Simulation 3, carry cold material further out, thus reducing the amount
that falls back.

It is interesting that Simulation 3 with a single sustained $50\Myr$ blast
yields a higher central temperature than Simulation 4 with two blasts of
$25\Myr$ duration. Two factors seem to contribute to this phenomenon. First,
after a burst lasting only $25\Myr$ much of the entrained cold material
falls straight back.  Second, the later blast in Simulation 4 has little
difficulty in linking up with the cavity blown by the first blast, which
continued to move outwards over the $25\Myr$ of quiescence between the
blasts. Consequently, the radii at which most of the cavity's energy is
deposited are larger than the radii heated by the second half of the
sustained blast in Simulation 3.  That is, a $25\Myr$ period of quiescence
is not long enough to make two blasts independent of one another.

\section{Conclusions}

A suite of three-dimension hydrodynamical simulations of the reheating of
intracluster gas by jets has been used to investigate the structural
stability of cooling flows that are episodically reheated by jets.

During a period of quiescence by the AGN, radiative cooling
rapidly increases the central concentration of the ICM. At some point the
rising central density of the ICM is expected to provoke an outburst by the
AGN. The precise instant at which the outburst comes is likely to vary from
outburst to outburst, and be correlated with the outburst's strength in the
sense that later outbursts that are fed by a more centrally concentrated ICM
are likely to be more powerful.
The more powerful a jet is, the further out it deposits its energy.

Simulations in which the time of the outburst is varied along with the
strength and/or duration of the outburst, suggest that cooling flows have a
tendency to structural stability in the following sense. On occasions when
there is an unusually long period of quiescence by the AGN and passive
cooling makes the intracluster gas unusually centrally concentrated
immediately before an outburst by the AGN, a larger fraction of the AGN's
energy is dissipated at small radii because, notwithstanding the increased
power or duration of the outburst, the jets are disrupted near the AGN by
the unusually dense ICM. Moreover, entrainment of cold gas by the jets is an
important mode of core heating, and the denser ICM increases the mass of
cold gas that is entrained.

Hence these simulations provide a framework for
understanding why the density profiles of cooling flows do not span a wide
range.

Integration of FR II sources into this framework is an important task for
the future. There is a tantalizing possibility that these intrinsically rare
sources are ones in which the stabilizing feedback loop that we have
described has broken down: above some critical power, too little energy is
deposited in the immediate vicinity of the black hole to throttle accretion
onto the hole, and a runaway growth in accretion rate, jet power and
physical source size may ensue. Both observations and simulations indicate
that powerful FR II sources deposit nearly all their energy outside the
cooling radius. Heating the ICM at such large radii can have little impact
on the black hole's accretion rate. Unfortunately, it is hard for
simulations to determine accurately what small fraction of a large FR II
power is dissipated near the centre, where the hole's accretion rate is
determined. The computational problem is akin to that involved in simulating
a core-collapse supernova, since here again an extremely small fraction of a
large energy budget is responsible for driving matter away from the
collapsing core.

It is worth noting that, at a given power, a narrower jet will deposit a
smaller fraction of its energy in the core because it will entrain less cold
gas. Hence FR II sources may be characterized by unusually narrow jets.

The existence of an effective feedback loop in cooling flows is easier to
understand if the black hole accretes directly from the hot ICM rather than
from an accretion disk fed with gas that has cooled catastrophically to
$<10^5\,$K. It is well known that the Bondi-Hoyle accretion rates of black
holes at the centres of cooling flows are more than sufficient to power the
cooling flow \citep{FabianCanizares88}. Moreover, the connection between
radio sources is well established 
\citep{Krolik}. This as yet unexplained connection would make sense if
energy released by accretion of hot gas is channelled into jets rather than
radiation, because elliptical galaxies are, almost by definition, systems in
which the interstellar medium is at the virial temperature, rather than made
up of cold, centrifugally supported gas. If the black holes in cooling flows
do indeed feed off virial-temperature gas, it follows that AGN effectively
prevent cooling of virial-temperature gas. This conclusion has radical
implications for the theory of galaxy formation and the origin of the galaxy
luminosity function \citep{binney04}. It also implies that the filaments of
cold gas that have been observed in several systems must be gas that fell in
cold \citep{SparksMG,NipotiB} rather than gas that has cooled from the X-ray
emitting phase.

\end{document}